# Simultaneous, Non-Contact Measurement of Liquid and Interfacial Thermal Properties via a Differential Square-Pulsed Source Method

Tao Chen, and Puqing Jiang*

School of Energy and Power Engineering, Huazhong University of Science and Technology, Wuhan, Hubei 430074, China

*E-mail: jpq2021@hust.edu.cn

**ABSTRACT:** Accurate characterization of heat transport across solid-liquid interfaces is essential for thermal management in micro- and nanoscale systems. Yet existing techniques often require prior knowledge of liquid properties, which complicates the simultaneous resolution of interfacial and bulk behaviors, and lose sensitivity once interfacial conductance exceeds 100 MW m$^{-2}$ K$^{-1}$. Here we present a differential square-pulsed source (DSPS) method that provides simultaneous, non-contact measurement of liquid thermal conductivity, volumetric heat capacity, and solid-liquid interfacial conductance without any predefined material parameters. Dual-frequency excitation combined with in-situ substrate referencing enables property extraction from multilayer structures, and numerical simulations show a typical uncertainty of about 8 % in interfacial conductance, confirming robustness. The protocol is validated for a wide spectrum of liquids, including oils, lubricants, aqueous electrolytes, and pure water, with excellent agreement with literature values for bulk properties. Analysis of the data set clarifies how vibrational-spectrum mismatch, ionic layering, and related interfacial phenomena govern heat transfer, and demonstrates that oleophilic hexadecyl silane modification of aluminum increases interfacial conductance by a factor of sixteen. The results reveal that conductance can be strongly tuned through surface wettability and chemical functionalization, offering direct guidelines for interface engineering. Because the approach is readily extendable to soft materials such as thermal-interface gels, it promises broad applicability in emerging interface-dominated thermal technologies.

## 1. INTRODUCTION

Thermal transport across solid-liquid interfaces plays a critical role in the performance of advanced thermal management systems,[1, 2] energy devices,[3, 4] and emerging nano- and microscale technologies.[5, 6] The solid-liquid interface is a key region where heat transfer resistance can significantly affect overall system performance. Accurate characterization of solid-liquid interfacial thermal conductance (ITC, $G$), along with the thermal properties of liquids such as thermal conductivity ($k_\text{f}$) and volumetric heat capacity ($C_\text{f}$), is essential for understanding and optimizing these processes. A representative example is form-stable phase-change materials (PCMs), where composite structuring with expanded-graphite scaffolds has been shown to increase the effective thermal conductivity of erythritol from ~0.6 to ~6 W m$^{-1}$ K$^{-1}$, underscoring the importance of reliable characterization of both bulk and interfacial thermal properties for functional materials.[7] In parallel, model-based and data-driven routes are increasingly used to analyze coupled fluid-solid heat transfer and to predict effective properties (e.g., 3-D fluid-solid coupling for industrial furnaces, and neural-regression models for CNT-nanofluid conductivity across multiple base fluids).[8, 9] However, the fidelity of such simulations and machine-learning (ML) frameworks ultimately hinges on high-quality experimental ground truth for $k_\text{f}$, $C_\text{f}$, and $G$. While bulk thermal properties of liquids are well-studied, the accurate measurement of ITC remains a challenging task, particularly for nanoscale systems.

The Kapitza length ($l_\text{k} = k_\text{f}/G$), introduced by Kapitza in 1941,[10] provides a practical measure of interfacial thermal resistance and is widely used to quantify heat transfer across solid-liquid boundaries. However, existing techniques often struggle to simultaneously measure the liquid's thermal properties and the ITC at the interface. ITC is particularly difficult to measure, as it is highly sensitive to both material properties and surface conditions, such as roughness, wettability, and molecular interactions. This challenge is compounded by the fact that many traditional techniques, such as steady-state methods,[11-15] transient hot-wire,[16, 17] transient plane source,[18-21] and laser flash analysis,[22-24] can only measure bulk liquids' thermal properties and lack the sensitivity to ITC.

Recent advances in methods like time-domain thermoreflectance (TDTR),[25-32] and frequency-domain thermoreflectance (FDTR)[33, 34] have made significant strides in measuring $k_\text{f}$ and solid-liquid ITC. For example, as a pioneering work, Cahill's group used TDTR to study the thermal transport characteristics at hydrophilic and hydrophobic Au-water and Al-water interfaces, revealing that the ITC of hydrophilic interfaces is 2-3 times larger than that of hydrophobic ones.[25] This result clearly demonstrates the decisive role of wettability in interfacial thermal coupling. More recently, Meng et al.[34] used FDTR to examine the ITC of Al with water, ethanol, and hexadecane, reaching conclusions consistent with the earlier study.

However, the inherently low thermal conductivity of liquids imposes two major limitations. First, these methods typically require the volumetric heat capacity of the liquid—obtained from other sources—as a known input. Second, they often lack sufficient sensitivity to solid-liquid

ITC,[29, 33] particularly when the interfacial conductance exceeds ~100 MW m$^{-2}$ K$^{-1}$. In some studies, such as Refs.[25, 33], even the thermal conductivity of the liquid is treated as a known parameter to ensure accurate determination of ITC. As a result, the uncertainties of these assumed parameters are fully propagated into the ITC calculation, inevitably leading to larger errors in the extracted ITC values. Although our previously developed square-pulsed source (SPS) method[35] is capable of simultaneously measuring the thermal conductivity and heat capacity of liquids, it also suffers from limited sensitivity to interfacial conductance due to the low thermal conductivity of liquids—similar to the challenges faced by TDTR and FDTR. Thus, there remains a significant gap in techniques that can simultaneously measure liquid thermal conductivity, volumetric heat capacity, and ITC .

To address these limitations, we introduce the differential square-pulsed source (DSPS) method, a novel technique that enables simultaneous, non-contact measurement of liquid thermal conductivity, volumetric heat capacity, and solid-liquid ITC. The DSPS method is unique in that it does not rely on prior knowledge of the liquid heat capacity, allowing for versatile and accurate measurements of both the bulk and interfacial thermal properties of liquids. A key advantage of the DSPS method is its differential measurement strategy, which effectively reduces the uncertainty in ITC extraction. This method is validated across a diverse range of liquids, including lubricants, oils, aqueous solutions, and pure water, with excellent agreement with literature values for thermal conductivity and volumetric heat capacity. Furthermore, the DSPS method offers new insights into the role of ITC at solid-liquid interfaces, particularly through the investigation of surface modifications that enhance oleophilicity or hydrophilicity, which significantly improve interfacial heat transfer. By enabling direct and simultaneous measurements of these critical parameters, the DSPS method provides a powerful tool for understanding and optimizing interfacial thermal transport in liquid and nanoscale systems.

## 2. METHODS

**2.1. Measurement Technique and Data Processing.** Figure 1a shows a schematic of the DSPS system. In this configuration, a function generator modulates a 458 nm pump laser with a square wave to induce periodic heating of the sample. Concurrently, a 785 nm probe laser monitors the periodic surface temperature fluctuations of the sample. When the induced temperature rise remains relatively modest (below approximately 10 K), the intensity variations of the reflected probe laser accurately follow the temperature dynamics. The reflected probe signal is converted into an electrical signal by a photodiode detector and subsequently analyzed using the periodic waveform analyzer (PWA) module of a digital lock-in amplifier, extracting the amplitude over a complete modulation period. In periodic thermoreflectance measurements, PWA offers narrow-band extraction with large dynamic reserve, enabling robust amplitude/phase readout; in contrast, oscilloscopes have been primarily used in earlier studies to monitor the noise level, while PWA further facilitated precise signal extraction.[36] The extracted amplitude data is then fitted to a heat transfer model to accurately determine the thermal properties of the sample. Detailed discussion on heat transfer model development is available in Supporting Information Section S1.

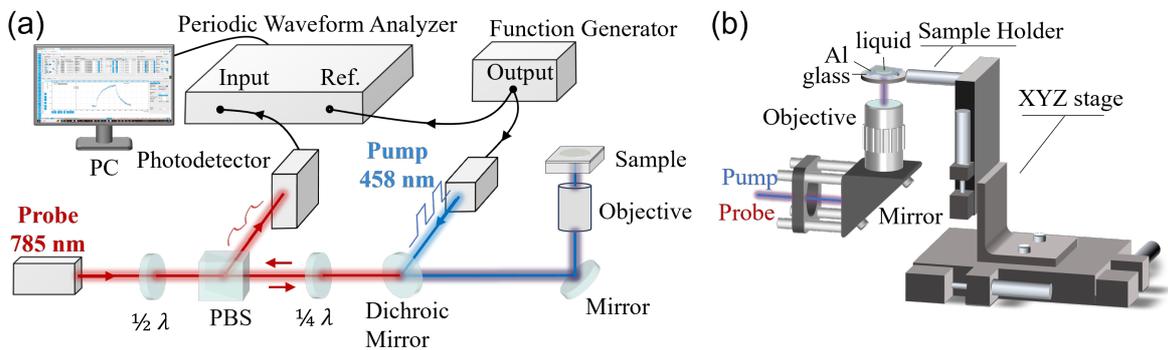

**Figure 1.** (a) Schematic of the differential square pulsed source (DSPS) setup for thermal measurement. (b) 3D optical layout showing pump and probe beams focused onto the sample.

The DSPS setup is similar to FDTR, but there are key differences. In FDTR, a beamsplitter is required at the pump beam output to measure the pump phase, and the optical paths of the pump and probe beams must be equal to determine the phase difference between the two beams as a function of the pump modulation frequency. In contrast, TDTR uses two pulsed lasers with a controlled time delay, applying a pulsed heating source and measuring the temperature evolution as a function of delay time. While DSPS shares some similarities with FDTR, its unique advantage lies in its high signal-to-noise (SNR) ratio and the ability to achieve a wide range of modulation frequencies, from 1 Hz to 10 MHz. The superior SNR of SPS originates from measuring the amplitude signal and employing PWA-based synchronous averaging, which ensure robust readout even at low modulation frequencies, where FDTR phase detection typically fails due to phase drift and near-zero response. This capability makes DSPS particularly effective for measuring materials with low in-plane thermal conductivity,[35] anisotropic structures exhibiting three-dimensional thermal conduction,[37] multilayer films,[38] and convective heat transfer coefficients.[39]

For measuring the thermal properties of liquid films, a microscope glass slide coated with an approximately 100 nm thick aluminum (Al) film was employed as the thermal transducer. As shown in Figure 1b, the transducer was positioned with the Al film facing upward. The pump and probe laser beams, directed from below, passed through the transparent glass substrate and were focused at the glass-Al interface. Liquid films were applied on the upper surface of the Al film. The low thermal conductivity of the glass promotes preferential heat flow toward the liquid side, enhancing measurement sensitivity to the thermal properties of the liquid film.

To minimize uncertainties associated with the Al layer, glass substrate, and the laser spot size, a differential measurement strategy was adopted. As illustrated in Figure 2a, identical measurements were conducted with and without the liquid film under the same experimental conditions. Since both measurements share the same substrate, transducer, and optical configurations, differential analysis effectively suppresses sensitivity to these common parameters, improving the accuracy of thermal property extraction for the liquid film.

Two sets of amplitude signals as a function of time over a full square-wave period are generated. The set corresponding to the liquid film case is denoted as $A(t)$, and the set for the no-liquid-film case is denoted as $A_0(t)$, with their respective maximum amplitudes $A_\mathrm{m}$ and $A_{0,\mathrm{m}}$, as illustrated in Figure 2b (time on the horizontal axis is normalized).

As shown in Figure 2c, these amplitude signals are then normalized into three dimensionless sets: $A(t)/A_\mathrm{m}$, $A_0(t)/A_{0,\mathrm{m}}$, and $A(t)/A_0(t)$. This normalization process effectively cancels out parameters such as laser power and thermal reflection coefficients, eliminating the need to know these parameters explicitly. The measurements are repeated using two distinct modulation frequencies: one in the hundreds of Hz and the other in the MHz range. This dual-frequency approach enables simultaneous determination of the liquid's thermal conductivity and volumetric heat capacity. Among these, the $A(t)/A_0(t)$ signal is primarily focused on the high-frequency response. This results in a total of five sets of normalized signals.

As illustrated in Figure 2d, the sample structure involves six unknown parameters, including the thermal conductivity $k_\mathrm{sub}$ and volumetric heat capacity $C_\mathrm{sub}$ of the substrate, the ITC $G_1$ between the metal transducer layer and the substrate, the thermal conductivity $k_\mathrm{f}$ and volumetric heat capacity $C_\mathrm{f}$ of the liquid film, and the ITC $G_2$ between the liquid film and the metal layer. Among these, $G_1$ exhibits extremely low sensitivity in all measurements and can therefore be treated as a known parameter without requiring fitting. As shown in Figure 2e, $k_\mathrm{sub}$ and $C_\mathrm{sub}$ are first simultaneously fitted using the $A_0(t)/A_{0,\mathrm{m}}$ signals at both frequencies. Then, the high-frequency signals $A(t)/A_\mathrm{m}$ and $A(t)/A_0(t)$ are used to simultaneously fit $k_\mathrm{f}C_\mathrm{f}$ and $G_2$. Once $k_f C_f$ is determined, the low-frequency $A(t)/A_m$ signal is used to decouple $k_f$ and $C_f$.

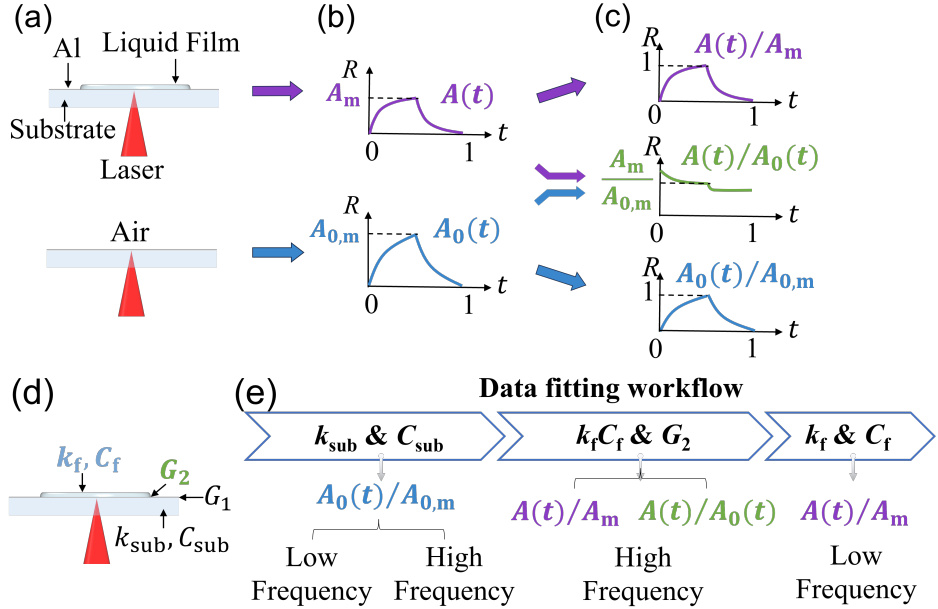

**Figure 2.** (a) Comparative DSPS measurements with and without a liquid film under identical laser conditions. (b) Raw amplitude signals $A(t)$ and $A_0(t)$ over a square-wave period. (c) Normalized signals $A(t)/A_m$, $A_0(t)/A_{0,m}$, and $A(t)/A_0(t)$. (d) Sample structure with six thermal parameters. (e) Sequential data fitting procedure to extract $k_{sub}$, $C_{sub}$, $G_2$, $k_f$, and $C_f$ from dual-frequency measurements.

**2.2. Uncertainty Estimation.** The uncertainty analysis of DSPS is based on a comprehensive error propagation formula to estimate the uncertainties of the parameters to be measured. This method was initially introduced by Yang and Schmidt for analyzing the uncertainty in FDTR experiments.[40] The formula is derived using least-squares fitting of the signal and accounts for not only the propagation of errors in input parameters but also the uncertainties from experimental noise and fitting quality.

In data processing, multiple parameters are extracted by simultaneously fitting $M$ sets of experimental signals using the least-squares regression method. Mathematically, this involves minimizing the product of the mean squared deviation (MSD) between each set of experimental signals and its corresponding model predictions:

$$J = \prod_{j=1}^{M} \frac{1}{N_j} \sum_{i=1}^{N_j} \left(g_j(\boldsymbol{U}, \boldsymbol{P}, t_i) - y_j(t_i)\right)^2 = \prod_{j=1}^{M} \text{MSD}_j \qquad (1)$$

Here, $j$ represents the $j$-th set of experimental signals, $y_j(t_i)$ is the signal measured at the i-th time point $t_i$, and $N_j$ is the total number of data points for the $j$-th signal. The function $g_j(\boldsymbol{U}, \boldsymbol{P}, t_i)$ represents the predicted value based on the thermal model, where $\boldsymbol{U}$ is the vector of unknown parameters to be fitted, containing $u$ parameters, and $\boldsymbol{P}$ is the vector of known input parameters, containing $p$ parameters.

In this study, $\boldsymbol{U} = (k_{\text{sub}}, C_{\text{sub}}, G_2, k_f, C_f)^T$. Although the signal $A_0(t)/A_{0,m}$ is independent of $G_2$, $k_f$, and $C_f$, setting $\boldsymbol{U}$ in this way is still valid, as it is sufficient to set the partial derivatives of $A_0(t)/A_{0,m}$ with respect to these parameters to zero. Additionally, $\boldsymbol{P} = (k_m, C_m, h_m, G_1, r_0)^T$, where the subscript m refers to the metal film, $r_0$ is the laser spot radius. The typical errors are 5% for $k_m$, 3% for $C_m$, 2% for $h_m$, and 2% for $r_0$.

In the case of optimal fitting, the partial derivatives of $J$ with respect to each element of $\boldsymbol{U}$ are zero. $\widehat{\boldsymbol{P}}$ represents a possible set of random values for the known parameters since there are uncertainties in these input parameters, while $\widehat{\boldsymbol{U}}$ corresponds to the fitted parameter set that minimizes the fitting error. The uncertainty of the unknown parameters can be revealed through the distribution of all possible $\widehat{\boldsymbol{U}}$ values. Based on this principle, through a series of derivations, the covariance matrix $\text{Var}[\widehat{\boldsymbol{U}}]$ is obtained, which has the form:

$$\text{Var}[\widehat{\boldsymbol{U}}] = \begin{bmatrix} \sigma_{U_1}^2 & \text{cov}[U_1, U_2] & \cdots \\ \text{cov}[U_2, U_1] & \sigma_{U_2}^2 & \cdots \\ \vdots & \vdots & \ddots \end{bmatrix} \quad (2)$$

Here, the elements on the main diagonal represent the variance of the normal distribution of the unknown parameters. The uncertainty of a parameter is taken as twice its standard deviation. A detailed derivation of $\text{Var}[\widehat{\boldsymbol{U}}]$ can be found in Supporting Information Section S2.

**2.3. Measurement Capability of DSPS.** To validate the broad-range measurement capabilities of the DSPS method for $k_f$, $C_f$, and $G_2$, extensive numerical experiments were performed. In total, 1,000 parameter sets were randomly selected, covering a broad range: thermal conductivity $k_f$ from 0.01 to 100 W m$^{-1}$ K$^{-1}$, volumetric heat capacity $C_f$ from 0.1 to 5 MJ m$^{-3}$ K$^{-1}$, and ITC $G_2$ from 1 to 500 MW m$^{-2}$ K$^{-1}$. Synthetic signals generated using a laser spot radius ($r_0$) of 15 $\mu$m and modulation frequencies of 500 Hz and 1 MHz were supplemented with realistic noise levels consistent with actual experimental conditions.

Figure 3a, Figure 3b, and Figure 3c show the fitting results and error bars for $k_f$, $C_f$, and $G_2$, respectively. The x-axis represents the designed values, while the y-axis denotes the corresponding fitted results. Two reference lines, indicating ±20% deviation from the designed values, are included in the plots, showing that most data points fall within this range.

Figure 3d presents violin plots and quartiles depicting the percentage deviation between the fitted and designed values. The results show that 100% of the $k_f$ values have deviations within 5%, 97% of the $C_f$ values are within 5%, and 95% of the $G_2$ values are within 20%, demonstrating the robustness of the DSPS method for simultaneously measuring $k_f$, $C_f$, and $G_2$ across a broad range.

Figure 3e further illustrates the error distribution for fitted results, with all $k_f$ and $C_f$ errors remaining below 10%, and typical errors averaging around 4.5%. The typical error for $G_2$ measurements is approximately 8%, although occasionally higher deviations occur, particularly under conditions of low $k_f$ and high $G_2$. Under these conditions, increasing the modulation frequency of the pump laser effectively mitigates $G_2$ measurement errors, enhancing overall measurement accuracy.

As shown in Figure 3f, error analyses of 1,000 cases for the Kapitza length $l_k$ (defined as $k_f/G_2$) under modulation frequencies of 1 MHz and 10 MHz reveal that using 10 MHz significantly reduces the error. When $l_k > 1$ nm, the majority of data points exhibit errors below 20%, indicating that the DSPS method is capable of reliably measuring $l_k$ for values greater than 1 nm.

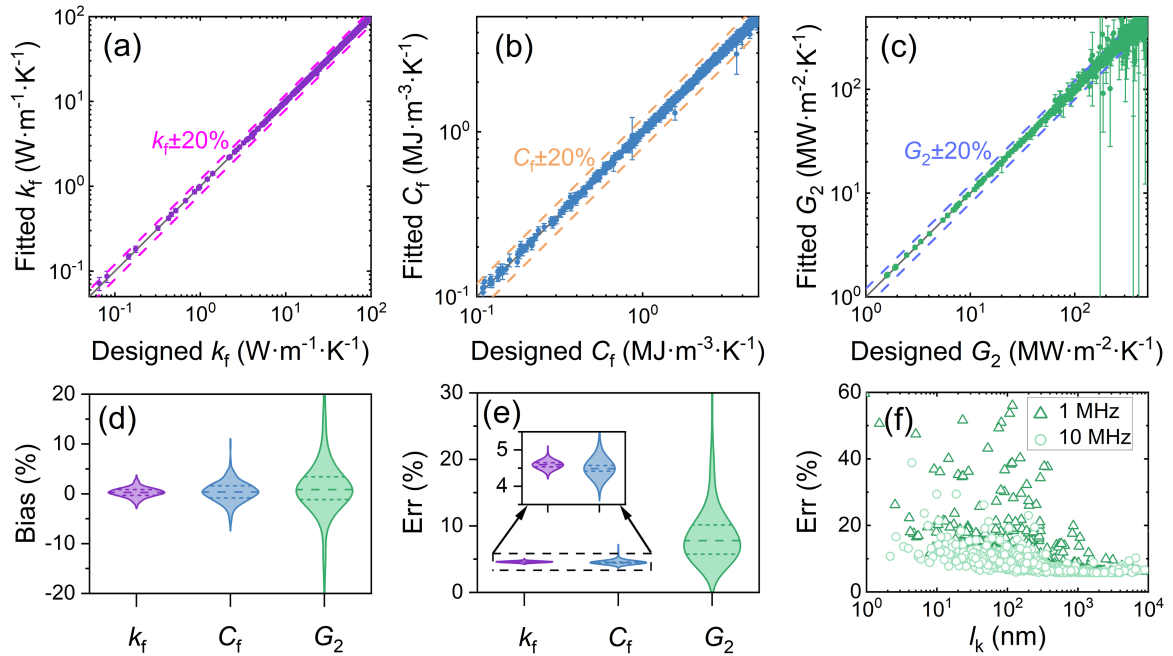

**Figure 3.** (a-c) Comparison of fitted and designed values of $k_f$, $C_f$, and $G_2$ for 1,000 synthetic data sets. (d) Fitting deviations showing >95 % of results within acceptable error bounds. (e) Error distributions with typical errors of ~4.5 % for $k_f$ and $C_f$, and ~8 % for $G_2$. (f) Error distribution of Kapitza length ($l_k = k_f/G_2$) from 1,000 cases under modulation frequencies of 1 MHz and 10 MHz, showing improved accuracy at 10 MHz and reliable measurement for $l_k > 1$ nm.

In summary, the DSPS method is theoretically capable of accurately measuring $k_f$, $C_f$, and $G_2$ over a wide range. It can, in principle, be applied to various soft materials such as liquids and thermal interface materials (TIMs). Such capability is particularly attractive for screening next-generation TIM gels, immersion-cooling fluids, and latent-heat storage media, where bulk-interface synergy dictates device reliability.

## 3. RESULTS AND DISCUSSION

### 3.1. Typical Case.
Figure 4 shows an example of measurement and data processing for the case of peanut oil. Square-pulsed source (SPS) measurements were conducted in two conditions: first, without a peanut oil droplet on the transducer surface, and then with a peanut oil droplet placed on the surface. Laser powers were kept constant across comparative measurements. The root-mean-square (RMS) laser spot size, defined as the average of the pump and probe laser spot sizes, was characterized using the spatial-domain thermoreflectance (SDTR) method as $r_0 = 11 \ \mu m$.[41] Two pump modulation frequencies, $f_L = 500$ Hz and $f_H = 1$ MHz, differing significantly in magnitude, were chosen to create distinct heat conduction modes.[35] Frequencies in the hundreds of kilohertz range were deliberately avoided to minimize sensitivity to the Al film thermal conductivity,[35] thereby reducing measurement uncertainty. Raw signals labeled $A_0^L$, $A_0^H$, $A^H$, and $A^L$ are shown in Figures 4a to 4d, where the subscript "0" denotes measurements without the liquid film, and the superscripts "L" and "H" indicate lower (500 Hz) and higher (1 MHz) modulation frequencies, respectively.

First, signals $A_0^L$ and $A_0^H$ were normalized, and the substrate's thermal conductivity ($k_{sub}$) and volumetric heat capacity ($C_{sub}$) were determined by minimizing deviations between normalized experimental signals ($A_{0,norm}^L$, $A_{0,norm}^H$) and simulated signals, as illustrated in Figures 4e and 4f. Corresponding sensitivity analyses (Figures 4g and 4h) defined sensitivity coefficients ($S_\alpha$) as $\partial(\ln R)/\partial(\ln \alpha)$, where $R$ is the measured signal and $\alpha$ the parameter analyzed. At $f_H = 1$ MHz, $k_{sub}$ and $C_{sub}$ appeared coupled as the product $k_{sub}C_{sub}$, yielding $k_{sub}C_{sub} = 1.5411$ MJ$^2$ m$^{-4}$ K$^{-2}$ s$^{-1}$. Due to the low sensitivity of Al/glass ITC ($G_1$), $G_1$ was treated as an input with a value of $150 \pm 75$ MW m$^{-2}$ K$^{-1}$.[26] Conversely, at $f_L = 500$ Hz, the parameters coupled as the ratio $k_{sub}/C_{sub}$, resulting in $k_{sub}/C_{sub} = 0.4688$ mm$^2$ s$^{-1}$; under this condition, assuming a natural convective heat transfer coefficient of 10 W m$^{-2}$ K$^{-1}$ at the Al surface led to a maximum signal variation of only 0.03%, confirming that convection effects can be neglected. Combining these outcomes provided $k_{sub} = 0.85 \pm 0.04$ W m$^{-1}$ K$^{-1}$ and $C_{sub} = 1.813 \pm 0.09$ MJ m$^{-3}$ K$^{-1}$. Uncertainties from Al film properties, laser spot size, and $G_1$ were propagated into these parameters, effectively limiting subsequent influence.

Next, signals $A^H$ and $A_0^H$ from high-frequency measurements (Figures 4b and 4c) were divided to suppress uncertainties from Al film properties, laser spot size, and $G_1$. The resulting processed signal and corresponding fitting curve are presented in Figure 4i. Sensitivity analysis (Figure 4j) shows that the ratio signal exhibits a sensitivity to $G_1$ lower than 0.005, indicating that the differential scheme effectively suppresses the propagation of $G_1$-related errors into the target parameters. In addition, the sensitivity to the Al-peanut oil ITC ($G_2$) gradually decreases with time, while the sensitivities to the liquid film's thermal conductivity ($k_f$) and volumetric heat capacity ($C_f$) increase markedly. As heat penetrates deeper into the liquid, the measured responses increasingly reflect intrinsic liquid properties. Negative sensitivity coefficients indicate enhanced heat diffusion with increasing parameter values, thus reducing thermal accumulation in the Al film. Crucially, $k_f$ and $C_f$ remained coupled as $k_f C_f$ but effectively decoupled from $G_2$, enabling simultaneous determination. Further normalization of signal $A^H$ (Figure 4k) exhibited opposing sensitivities for $k_f C_f$ and $G_2$ (Figure 4l). Simultaneous fitting of signals $A^H/A_0^H$ and $A_{norm}^H$ yielded $k_f C_f = 0.315$ MJ$^2$ m$^{-4}$ K$^{-2}$ s$^{-1}$ and $G_2 = 16 \pm 2$ MW m$^{-2}$ K$^{-1}$.

In the final analysis step, signal $A^L$ was normalized (Figure 4m), effectively decoupling sensitivities between $k_f$ and $C_f$ (Figure 4n). Holding $k_f C_f$ constant and adjusting $C_f$ to fit experimental data, optimal results were obtained: $k_f = 0.17 \pm 0.01$ W m$^{-1}$ K$^{-1}$ and $C_f = 1.85 \pm 0.09$ MJ m$^{-3}$ K$^{-1}$. These measurements correspond to a Kapitza length ($l_k$) at the Al-peanut oil interface of approximately 10.6 nm.

At the lower modulation frequency (500 Hz), the thermal penetration depth calculated as $d_p = \sqrt{k_f/(\pi f_L C_f)}$ is approximately 8 $\mu$m. To ensure that the heat does not fully penetrate the liquid film, the film thickness should exceed three times this penetration depth,[41] setting a minimum thickness requirement of about 24 $\mu$m for accurate measurements. In this study, the liquid film used was sufficiently thick, not less than 1 mm, which is far greater than this threshold.

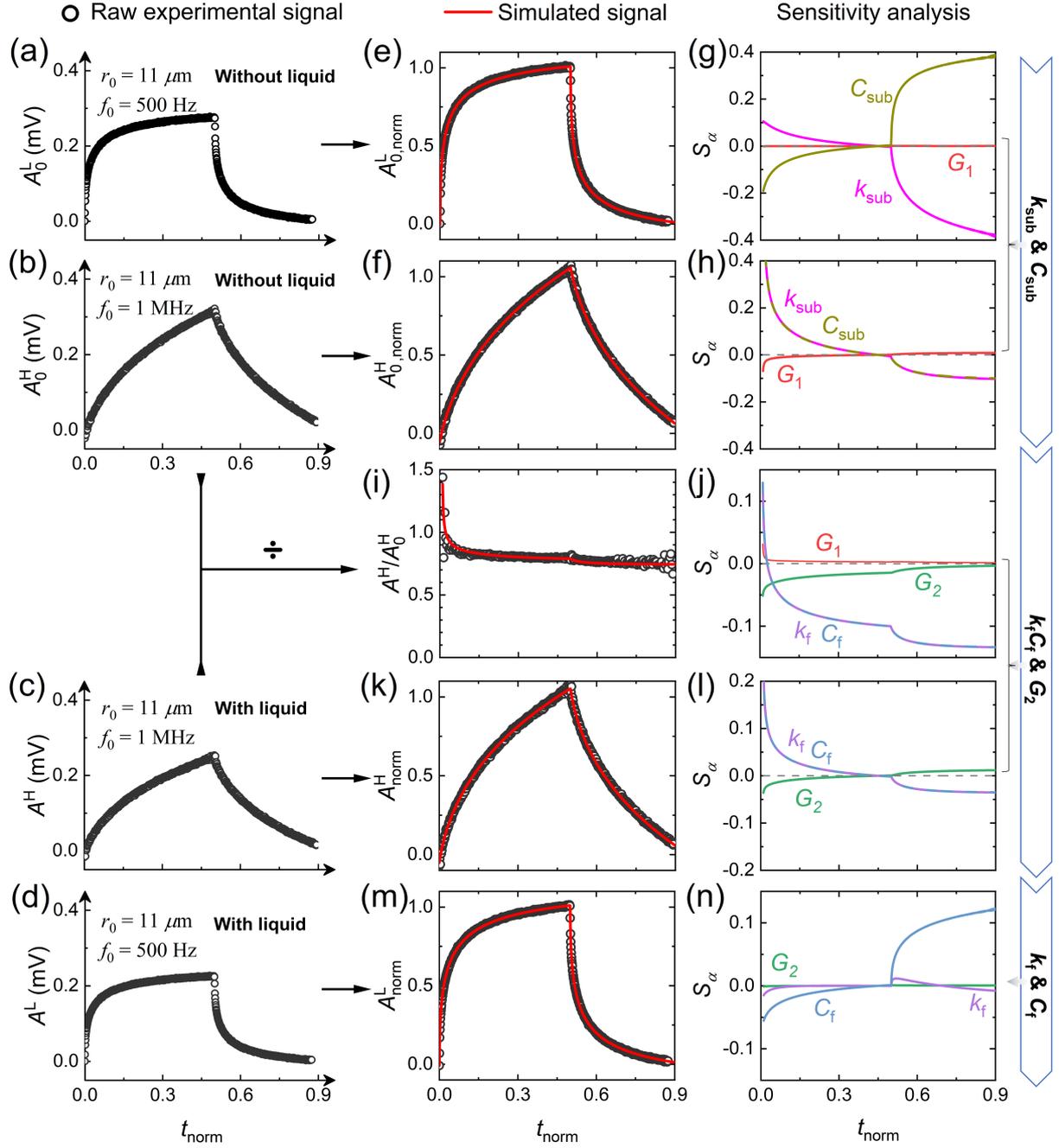

**Figure 4.** (a–d) Raw amplitude signals at 500 Hz and 1 MHz before and after adding the liquid film. (e–h) Fitting and sensitivity analysis of normalized signals to extract substrate thermal properties. (i, j) High-frequency signal ratio $A^H/A_0^H$ and its sensitivity to $k_fC_f$ and $G_2$. (k, l) Further normalization for decoupling $k_fC_f$ and $G_2$. (m, n) Low-frequency normalization and sensitivity enabling separation of $k_f$ and $C_f$.

**3.2. Results of Measurements on Various Liquids.** We applied the DSPS method to measure thermal properties of various liquid films at room temperature, including WD-40 lubricant, a tributyl phosphate (TBP)-dodecane mixture (volume ratio 3:7), peanut oil, ethanol, a 25 wt% NaCl aqueous solution, and pure water. Figure 5a and 5b present comparisons between experimentally measured thermal conductivities and volumetric heat capacities and their corresponding literature values, showing excellent agreement and thus validating the reliability of our measurement approach. The measured thermal conductivities and Kapitza lengths ($l_k$) at the liquid-Al interfaces are summarized in Figure 5c, with specific values listed in Table 1.

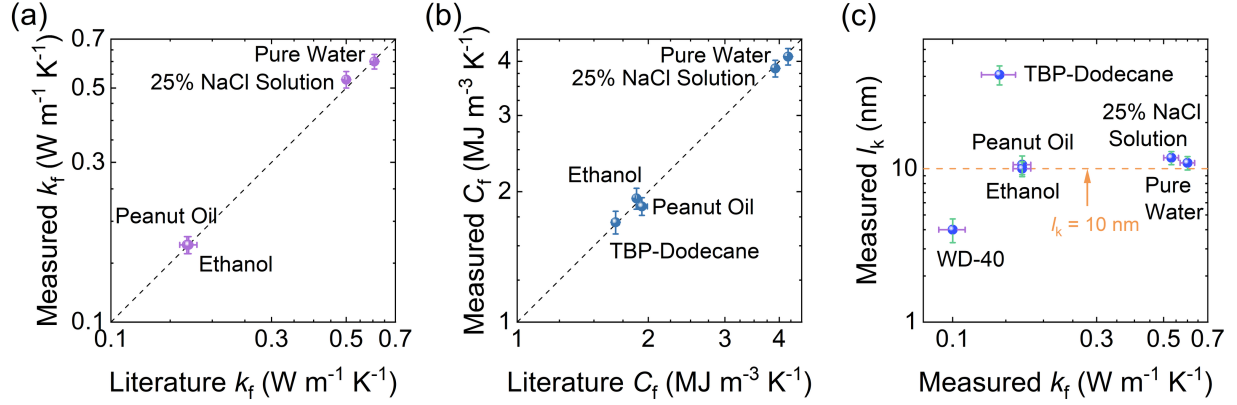

**Figure 5.** (a) Thermal conductivity and (b) volumetric heat capacity of various liquids measured in this work compared with literature data. (c) Kapitza length at Al/liquid interfaces as a function of liquid thermal conductivity.

**Table 1.** Measured $k_f$, $l_k$, and $C_f$ of various liquids with uncertainties, along with literature values for $k_f$ and $C_f$.

| | Measured | | | Literature | |
| --- | --- | --- | --- | --- | --- |
| Liquid | $k_f$ [W m$^{-1}$ K$^{-1}$] | $l_k$ [nm] | $C_f$ [MJ m$^{-3}$ K$^{-1}$] | $k_f$ [W m$^{-1}$ K$^{-1}$] | $C_f$ [MJ m$^{-3}$ K$^{-1}$] |
| WD-40 | 0.1±0.01 | 4±0.7 | 1.8±0.125 | - | - |
| TBP-dodecane | 0.14±0.02 | 41±5.8 | 1.7±0.094 | - | 1.682±0.034[42, 43] |
| peanut oil | 0.17±0.01 | 10.6±1.5 | 1.85±0.086 | 0.17±0.003[44] | 1.933±0.058[45] |
| ethanol | 0.17±0.01 | 10±1.1 | 1.93±0.107 | 0.17±0.01[29] | 1.879±0.012[46] |
| 25% NaCl Solution | 0.53±0.03 | 11.8±1.2 | 3.85±0.17 | 0.5±0.005[47] | 3.915±0.013[47] |
| pure water | 0.6±0.03 | 10.9±1.1 | 4.1±0.18 | 0.6 ± 0.004[48] | 4.189±0.010[49] |

Our results indicate that peanut oil, ethanol, and pure water exhibit similar Kapitza lengths of approximately 10 nm. This consistency is likely attributable to similar interfacial interactions between these liquids and the aluminum surface, along with the dominant role played by surface roughness in determining thermal boundary resistance.

However, certain liquids markedly deviate from this general trend. For instance, the 25 wt% NaCl aqueous solution exhibits a slightly higher Kapitza length than the pure water-Al interface.

This increase is mainly due to the solution's elevated ionic strength, which induces overscreening at the aluminum surface and forms a thicker Stern layer. Simultaneously, penetration by $Na^+$ and $Cl^-$ ions disrupts the first two layers of interfacial water, weakening water-Al phonon coupling. The combined effect significantly hampers heat transport across the interface.[50, 51]

Conversely, the WD-40 lubricant displays an unusually small Kapitza length of about 4 nm. This significant reduction arises from its unique composition, primarily consisting of hydrogenated light distillates and carbon dioxide. The low viscosity and high fluidity of these hydrogenated distillates enhance interfacial wettability, creating a more uniform liquid-solid contact layer. Additionally, dissolved carbon dioxide further reduces surface tension, thus promoting stronger interfacial coupling. Collectively, these factors enhance the solid-liquid adhesion and sharply lower the thermal boundary resistance.[28, 52]

In sharp contrast, the TBP-dodecane mixture exhibits a notably large Kapitza length of approximately 41 nm. This substantial interfacial resistance is explained by the significant mismatch in vibrational spectra across the interface. Molecular vibrations within the liquid occur at markedly higher frequencies, specifically the P=O stretching mode of TBP around 1278 cm$^{-1}$ (≈38 THz)[53] and the C-H stretching modes of n-dodecane around 2850 cm$^{-1}$ (≈85 THz),[54] whereas aluminum's phonon spectrum extends only up to approximately 9-10 THz.[55, 56] The pronounced vibrational mismatch between the liquid and aluminum severely impedes phonon transmission, resulting in a significantly enhanced thermal boundary resistance.[57, 58]

**3.3. Effect of Surface Modification on the Interface.** Figure 5c shows that the Kapitza length at the Al/TBP-dodecane interface is significantly larger than those of the other tested liquids, indicating substantially higher interfacial thermal resistance. This limitation, along with the broader goal of improving heat transfer at liquid-solid interfaces, motivated us to investigate surface modification strategies for enhancing interfacial conductance. To this end, we applied two distinct treatments to the aluminium surface: immersion in hexadecyltrimethoxysilane (HDTMS) to produce an oleophilic surface, and oxygen plasma exposure using a plasma asher to create a hydrophilic surface. We then evaluated the effects of these modifications using TBP-dodecane and water as representative liquids with contrasting polarity and practical relevance in thermal management applications.

Figure 6a shows that the water contact angle on untreated Al is 50°, indicating moderate wettability. After surface modification with HDTMS, the contact angle increased significantly to 104° (Figure 6b), even one month after treatment. This increase is attributed to the formation of a dense hydrophobic monolayer on the Al surface by the long-chain alkyl groups of HDTMS, effectively lowering surface energy and enhancing water repellency. Additionally, this surface treatment significantly improved the oleophilicity, enhancing wetting behavior for nonpolar liquids like TBP-dodecane. Conversely, Figure 6c presents the contact angle of 36° for Al treated with oxygen plasma, measured 34 hours after treatment, demonstrating sustained improvement in surface hydrophilicity. This improvement results from oxygen plasma effectively removing organic contaminants and introducing polar oxygen-containing functional groups such as hydroxyl and carbonyl groups, markedly enhancing the hydrophilicity of the surface.[59]

Subsequent DSPS measurements of the oleophilic Al/TBP-dodecane interface were performed at a pump modulation frequency of 9 MHz, selected for its shallow thermal penetration depth optimal for interfacial studies. Figure 6d displays amplitude signals with and without TBP-dodecane, and their ratio is shown in Figure 6e. With previously measured thermal properties of TBP-dodecane, the ITC was the only fitted parameter, yielding $G_2 = 56.7 \pm 14.0$ MW m$^{-2}$ K$^{-1}$, corresponding to $l_k = 2.47 \pm 0.64$ nm—a 94% reduction compared to the untreated interface, indicating significantly decreased thermal resistance. Similarly, measurements at the hydrophilic Al/water interface using the same frequency yielded amplitude signals in Figure 6f, with the signal ratio shown in Figure 6g. Fitting the amplitude ratio resulted in $G_2 = 147 \pm 23$ MW m$^{-2}$ K$^{-1}$ ($l_k = 4.08 \pm 0.68$ nm), representing a 63% reduction compared to the untreated surface, further confirming the notable improvement in ITC.

These results indicate that both oleophilic and hydrophilic surface chemical modifications can significantly reduce thermal resistance at solid-liquid interfaces, demonstrating that higher wettability generally leads to stronger ITC, consistent with the findings reported in Refs. [60, 61]. This strategy offers a clear and practical pathway for optimizing liquid-solid interfaces in applications such as thermal interface materials, immersion cooling technologies, and energy conversion systems.

To our knowledge, no prior study has directly quantified the Kapitza conductance across a metal-oil interface modified by HDTMS. Even the most effective self-assembled monolayer (SAM) treatments on metal-liquid interfaces have reported conductance enhancements by a factor of only 2 to 5.[25, 29] In contrast, our oleophilic modification results in an approximately 16-fold increase in ITC, highlighting the exceptional potential of HDTMS in engineering high-efficiency solid-liquid thermal interfaces.

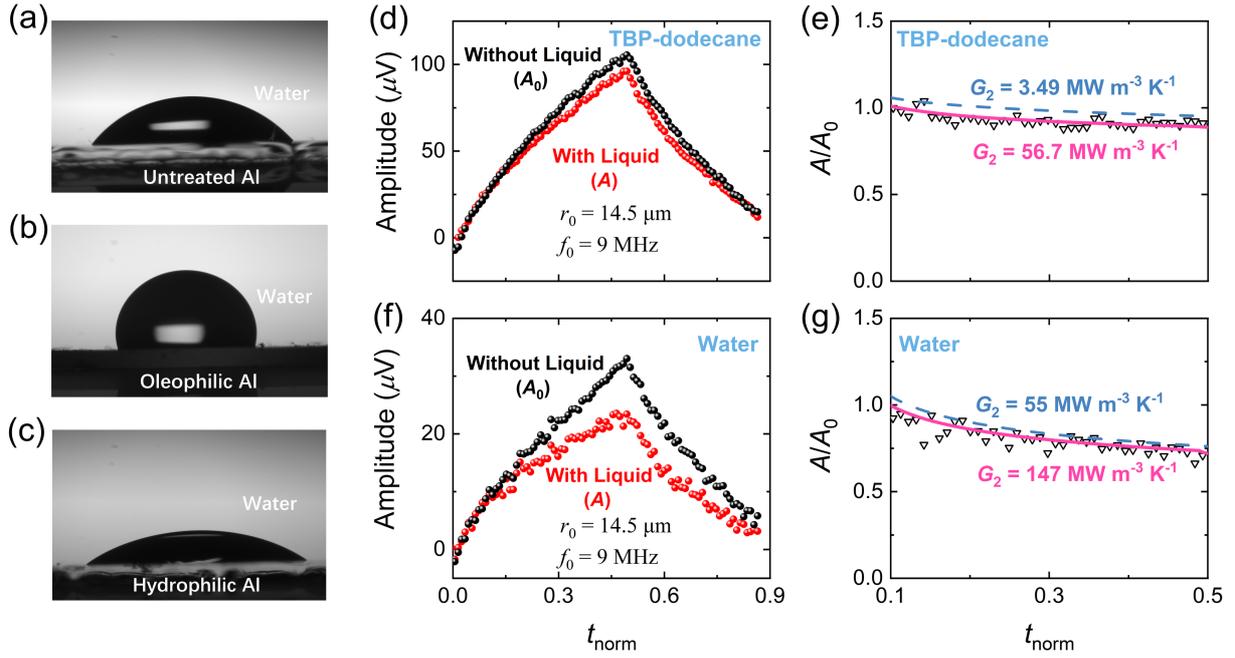

**Figure 6.** Characterizaiton of Al surface wettability and solid-liquid ITCs. (a-c) Water contact angle measurements on (a) untreated, (b) oleophilic, and (c) hydrophilic Al surfaces. (d, f) 9 MHz SPS measurement signals for Al/liquid interfaces with and without liquid present for (d) oleophilic Al/TBP-dodecane and (f) hydrophilic Al/water. (e, g) Ratios of the amplitude signals from (d) and (f), respectively. The best fits to the experimental data (red solid lines) yield an interfacial conductance of (e) $G_2 = 56.7$ MW m$^{-2}$ K$^{-1}$ for the oleophilic surface and (g) $G_2 = 147$ MW m$^{-2}$ K$^{-1}$ for the hydrophilic surface. For comparison, simulations using the untreated surface's conductance values (blue dashed lines: (e) $G_2 = 3.49$ MW m$^{-2}$ K$^{-1}$, (g) $G_2 = 55$ MW m$^{-2}$K$^{-1}$) deviate markedly from the data, demonstrating the significant enhancement of ITC achieved by surface treatment.

**3.4. Implications and Outlook.** The present dataset clarifies three actionable levers for engineering high-conductance solid-liquid interfaces: (i) vibrational matching, where minimizing spectral gaps between the liquid molecular modes and the solid phonon bandwidth reduces the Kapitza length; (ii) interfacial structure and charge, where ionic layering and overscreening can suppress energy transmission by disrupting the first hydration layers; and (iii) wettability and chemistry, where oleophilic or hydrophilic functionalization strengthens interfacial coupling, as evidenced by the approximately sixteen-fold enhancement for HDTMS on Al with TBP-dodecane. Practically, DSPS provides a rapid, non-contact screen for candidate fluids such as immersion coolants and electrolytes, as well as soft materials including TIM gels and PCM composites, with all three key parameters obtained in a single protocol. Our next work will apply this method to investigate the properties of TIMs under different pressures. In this scenario, the heat capacity becomes particularly important, further highlighting the advantage of DSPS in simultaneously measuring all three parameters.

## 4. CONCLUSIONS

In summary, we employ the differential square-pulsed source (DSPS) method to simultaneously measure the thermal conductivity, volumetric heat capacity, and solid-liquid interfacial thermal conductance of various liquid films, and the experimental results closely match literature values, validating the accuracy and robustness of the DSPS technique. Beyond benchmarking bulk properties, DSPS reveals how interfacial physics (vibrational mismatch, ionic effects) and surface chemistry (oleophilic and hydrophilic treatments) regulate interfacial conductance, offering valuable guidance for designing and optimizing solid-liquid interfaces and measurement approaches for liquids at the micro- and nanoscale. Looking forward, DSPS can be combined with in-situ surface characterization and controlled frequency/spot-size tuning to map the thermal properties of complex liquid-solid systems and to enable time-resolved studies during phase change.

## ASSOCIATED CONTENT

### Supporting Information

The Supporting Information is available free of charge at https://pubs.acs.org/doi/...

　　Details on the heat transfer model and the derivation of uncertainty calculations (PDF)

## AUTHOR INFORMATION

### Corresponding Author

**Puqing Jiang** - *School of Energy and Power Engineering, Huazhong University of Science and Technology, Wuhan, Hubei 430074, China;* orcid.org/0000-0002-1942-9259

　　E-mail: jpq2021@hust.edu.cn

### Author

**Tao Chen** - *School of Energy and Power Engineering, Huazhong University of Science and Technology, Wuhan, Hubei 430074, China;* orcid.org/0009-0002-5850-0959

Complete contact information is available at: https://pubs.acs.org/doi/...

### Notes

The authors declare no competing financial interests.

**ACKNOWLEDGMENTS**

This research was supported by the National Natural Science Foundation of China (NSFC) under Grant No. 52376058.

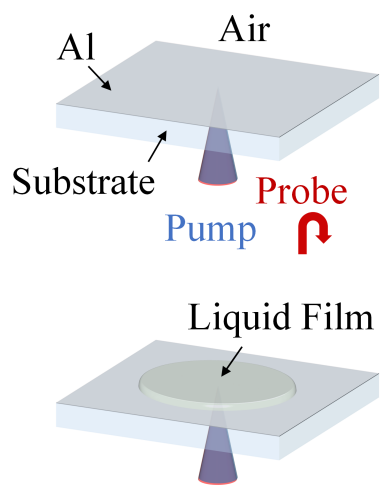
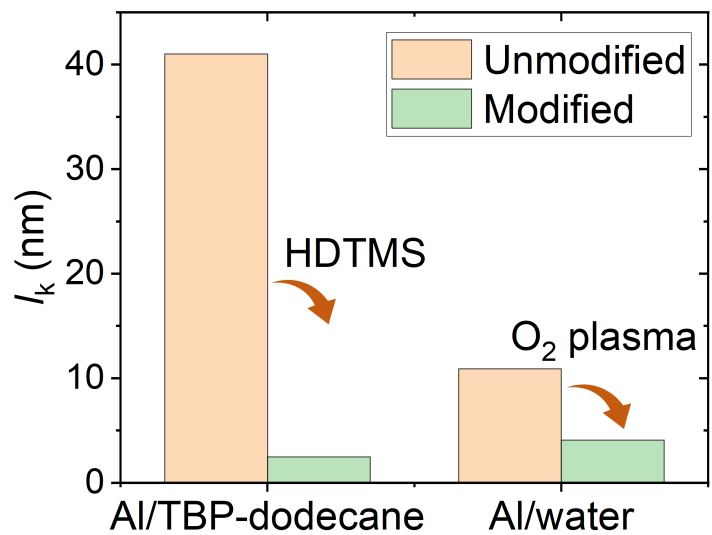